\documentclass[iop]{emulateapj-rtx4}
\usepackage{natbib}

\usepackage{comment}
\usepackage{lscape}
\usepackage{amsmath}
\usepackage{hyperref,breakurl}

\newcommand{\be}{\begin{equation}}
\newcommand{\ee}{\end{equation}}

\newcommand{\gray}{$\gamma$-ray}

\newcommand{\hi}{H~{\sc i}}

\newcommand{\fermilat}{{\it Fermi}-LAT}

\newcommand{\fermi}{{\it Fermi}}

\shorttitle{COSMIC-RAY PROPAGATION IN 30~DOR \& THE LMC}
\shortauthors{MURPHY~ET~AL.}
\slugcomment{Submitted to ApJ January 24, 2012; Accepted March 7, 2012}

\begin{document}
\title{Characterizing Cosmic-Ray Propagation in Massive Star-Forming Regions: The Case of 30~Doradus and the Large Magellanic Cloud}

\author{
E.\,J.\,Murphy$^{1}$, 
T.\,A.\,Porter$^{2,3}$,
I.\,V.\,Moskalenko$^{2,3}$, 
G.\,Helou$^{4}$, and
A.\,W.\,Strong$^{5}$
}

\affil{
$^{1}$\,Observatories of the Carnegie Institution for Science, 813 Santa Barbara Street, Pasadena, CA 91101, USA; emurphy@obs.carnegiescience.edu\\
$^{2}$\,Hansen Experimental Physics Laboratory, Stanford University, Stanford, CA 94305; tporter@stanford.edu, imos@stanford.edu\\ 
$^{3}$\,Kavli Institute for Particle Astrophysics and Cosmology, Stanford University, Stanford, CA 94305, USA\\
$^{4}$\,California Institute of Technology, MC 314-6, Pasadena, CA 91125, USA; gxh@ipac.caltech.edu\\
$^{5}$\,Max-Planck-Institut f\"ur extraterrestrische Physik, Postfach 1312, D-85741 Garching, Germany; aws@mpe.mpg.de
}

\begin{abstract}  
 
Using infrared, radio, and \gray{} data,
we investigate the propagation characteristics of cosmic-ray (CR) 
electrons and nuclei in the 30~Doradus (30\,Dor) star-forming region in 
the Large Magellanic Cloud (LMC)
using a phenomenological model based on the 
radio--far-infrared correlation within galaxies.
Employing a correlation analysis, 
we derive an average propagation length of $\sim 100-140$~pc 
for $\sim3$~GeV CR electrons resident in 30~Dor from consideration
of the radio and infrared data.
Assuming that the observed \gray{} emission towards 30~Dor is associated 
with the star-forming region, and applying the same methodology to the 
infrared and \gray{} data, we estimate a $\sim20$~GeV 
propagation length of $200-320$~pc for the CR nuclei.  
This is approximately twice as large as for $\sim3$~GeV CR electrons, 
corresponding to a spatial diffusion coefficient that is $\sim4$ times higher, 
scaling as $(R/{\rm GV})^{\delta}$ with $\delta \approx 0.7-0.8$ depending on
the smearing kernel used in the correlation analysis.
This value is in agreement with the results found by extending the 
correlation analysis to include $\sim70$~GeV CR nuclei traced by 
the $3-10$~GeV \gray{} data ($\delta \approx 0.66\pm0.23$).
Using the mean age of the stellar populations in 30~Dor and the 
results from our correlation analysis, we estimate a diffusion 
coefficient $D_{R} \approx 0.9-1.0 \times10^{27} (R/{\rm GV})^{0.7}$\,cm$^2$\,s$^{-1}$.
We compare the values of the CR electron propagation length and surface 
brightness for 30~Dor and the LMC as a whole with those of entire disk 
galaxies.   
We find that the trend of decreasing average CR propagation distance with 
increasing disk-averaged star formation activity holds for the LMC, and 
extends down to single star-forming regions, at least for the case of 30~Dor.
\end{abstract}

\section{Introduction}
Cosmic rays (CRs) are a dynamically important component of the interstellar 
medium (ISM) in galaxies.
Yet their role in shaping galaxy evolution is currently unclear due to the 
difficulties 
in characterizing their origin and 
propagation in the ISM \citep[e.g.,][]{as07}.
The energy density of CRs is comparable to that of magnetic fields, as well 
as the radiation fields and turbulent motions of the interstellar gas in 
galaxies.
Together, CRs and large-scale magnetic fields comprise a relativistic plasma 
whose interactions with the interstellar gas shape the overall chemistry 
and heating of the ISM \citep[e.g.,][]{bc90, kmf01, cox05}, and may 
also play a significant role regulating star-formation 
processes \citep[e.g.,][]{socrates08,ppp10,ppp11}.

The diffuse emissions from the radio to high-energy $\gamma$ rays ($> 100$ MeV),  
produced by various interactions between CRs and the interstellar gas, 
radiation and, magnetic fields, are the best way to characterize 
the physics of CRs throughout most of the Milky Way, and other galaxies.  
For example, the CR lepton component can be probed by radio, X-ray, 
and \gray{} observations.  
The radio emission arising from the injection of CR electrons from supernova remnants trace these particles as they propagate through large-scale magnetic fields and lose energy from synchrotron radiation.  
Similarly, the same CR electrons inverse Compton (IC) scatter off the 
interstellar radiation field (ISRF) and cosmic microwave background, 
yielding observable emissions from X-ray to \gray{} energies \citep{tp08}.  
High-energy $\gamma$ rays are particularly useful because this energy range gives 
access to the dominant hadronic component in CRs via the observation 
of $\gamma$ rays from the decay of neutral pions produced by inelastic collisions 
between CR nuclei and the interstellar gas \citep{pf63}.

\begin{figure*}[th]
\begin{center}
\epsscale{1.17}
\plottwo{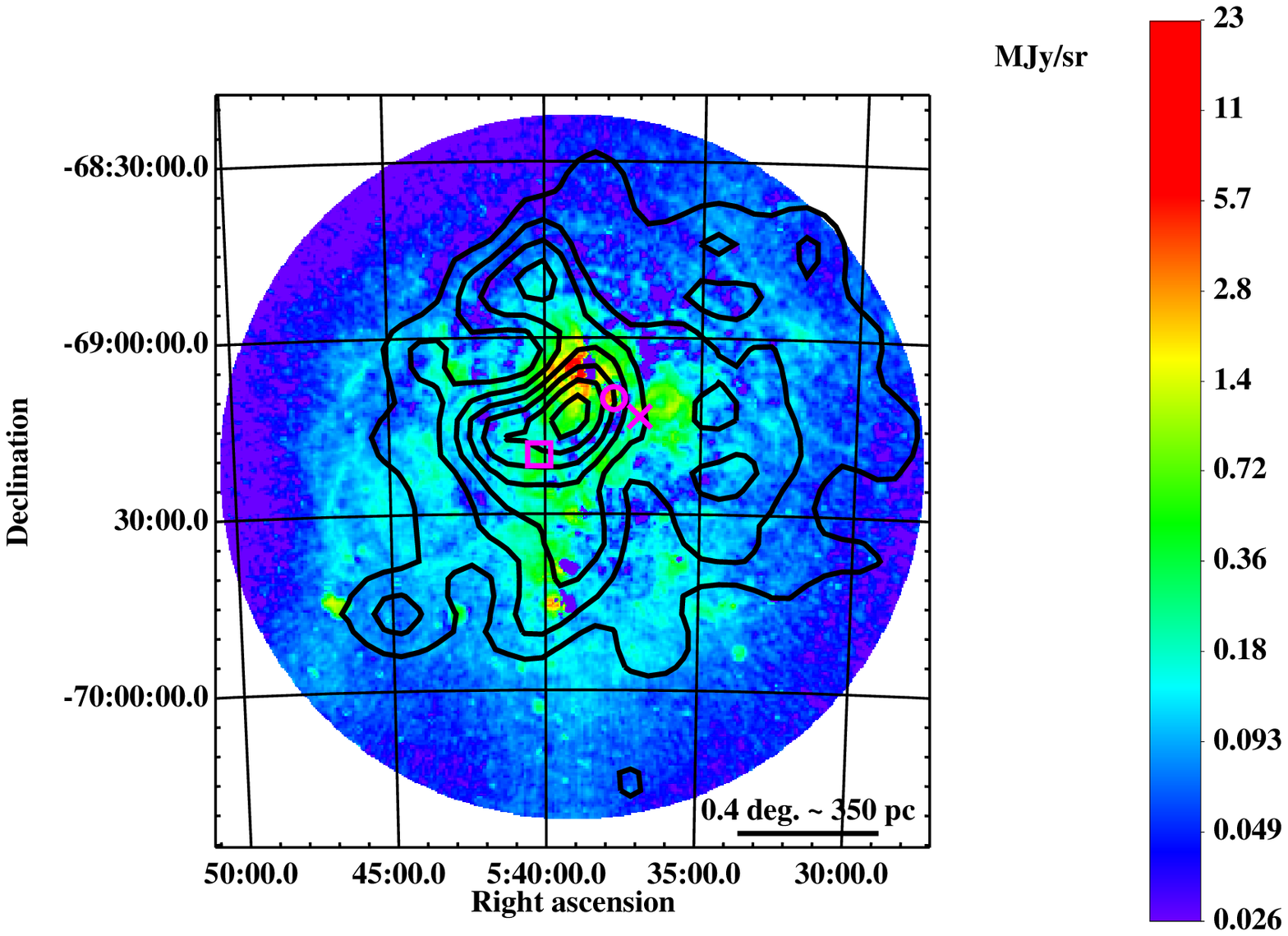}{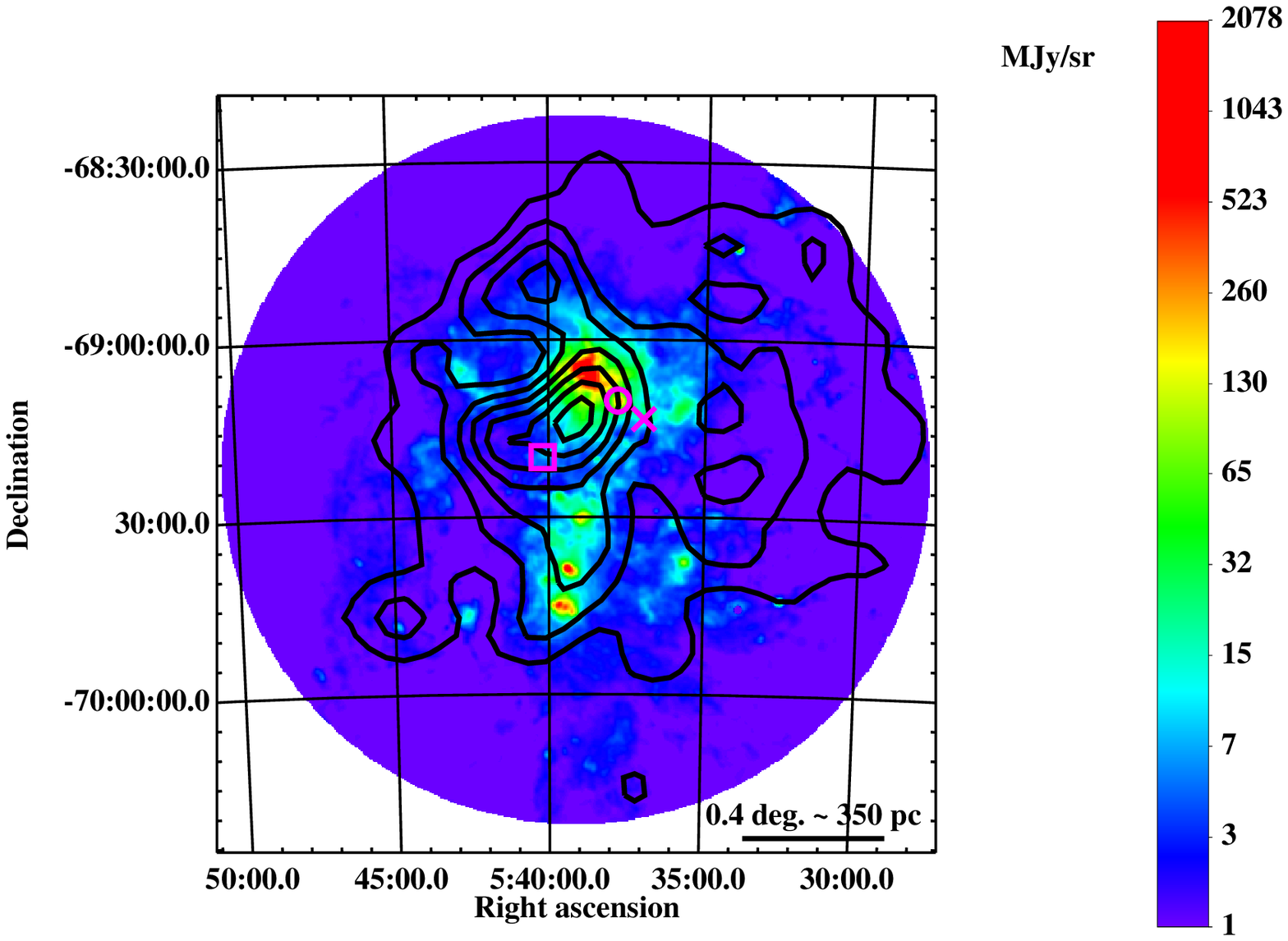}
\end{center}
\caption{
The point-source subtracted 1.4~GHz (left) and 24~$\mu$m (right) maps 
surrounding 30~Dor with residual $1-3$~GeV \gray{} contours starting 
at $0.5 \times 10^6$ counts sr$^{-1}$ (roughly $\sim$1 count per $0.1\degr \times 0.1\degr$ pixel) with linear intervals of $0.326 \times 10^6$ counts sr$^{-1}$ overlaid.
The 1.4~GHz map has been corrected for free-free emission.  
The region shown is centered at RA~$05^{\rm h}~39^{\rm m}~07^{\rm s}$, 
Dec.$-69^\circ 22\arcmin 02\arcsec$ (J2000) 
and has a radius of 1$^\circ$ ($\approx875$~pc).
The residuals between the radio/\gray{} and infrared maps were 
calculated over this region ($\S$\ref{sec-imsm}), 
as well the minimum energy magnetic field ($\S$\ref{sec-datradcre}), and the 
radiation field energy density ($\S$\ref{sec-imsm}).   
In addition, we show the locations of two Crab-like 
pulsars, PSR~J0540$-$6919 (square) and PSR~J0537$-$6910 (circle), 
along with a known background X-ray source RX~J0536.9$-$6913 (`X') that are 
coincident with the line-of-sight towards 30~Dor on the 1.4~GHz 
and 24~$\mu$m maps.  
\label{fig-1}}
\end{figure*}

Externally viewed galaxies have the advantage of mitigating line-of-sight
confusion that hampers interpretation of the diffuse emissions of the Milky Way.
While such observations can be used 
to determine the present distribution of CRs, understanding their 
propagation history remains difficult without knowledge of the 
initial distribution of CR sources.  
Using the tight, empirical correlation between the far-infrared (FIR) 
and (predominantly) non-thermal radio continuum emission from 
galaxies~\citep{de85,gxh85}, a number of studies have 
attempted to characterize the propagation of CR electrons in 
external galaxies \citep[e.g.,][]{mh98,ejm06a,ejm08}.   
The underlying physics relating the radio and FIR emission from 
galaxies lies in the process of massive star formation: young massive stars are 
the dominant sources of dust heating, and end their lives as supernovae 
whose remnants presumably accelerate and inject CR electrons into the ISM 
where they produce diffuse synchrotron emission.  
It was hypothesized that the radio image of a galaxy should resemble a 
smoothed version of its infrared image, because 
the mean free path of dust-heating photons ($\sim100$~pc) is significantly 
shorter than the typical diffusion length of CR 
electrons ($\sim1-2$~kpc) \citep{bh90}.  
This phenomenology is supported by studies within nearby galaxies 
at kpc \citep[e.g.,][]{mh98} and few hundred pc \citep[e.g.,][]{ejm06a} 
scales, as well as on scales of $\ga 50$~pc in the Large
Magellanic Cloud \citep[LMC,][]{ah06}.   
Investigations on sub-kpc scales within a sample of nearby spirals 
have even shown a dependence of the typical propagation length of CR 
electrons on star-formation activity arising from the predominant youth of 
CR electron populations in galaxies with enhanced disk-averaged star 
formation rates \citep{ejm06b, ejm08}.  

Recently, using 11 months of data, the LMC was detected by the \fermilat\ \citep{aa10-lmc} at high significance with the resolved \gray{} emission showing very little correlation with the large-scale distribution of the gas column density.  
If CRs freely diffuse in the ISM 
of the LMC, as they appear to in the Milky Way, the \gray{} emission
should be correlated with the distribution of gas, which is predominantly 
neutral hydrogen and helium in the LMC \citep{lss03}. 
Instead, the observed \gray{} emission for the LMC is more strongly 
correlated with tracers of massive star-forming regions.
Employing the multi-frequency diffuse emissions 
from radio to \gray{} energies, we can investigate for the first time the 
propagation of both CR electrons and nuclei 
associated with a star-forming region in an externally resolved galaxy, 
30~Dor in the LMC.


\section{Data and Analysis}
We compiled archival infrared and radio data, together with 32 months of 
\gray{} data from the \fermilat\ for our analysis.
We assume a distance 
to 30~Dor of 50~kpc throughout this analysis\footnote[1]{Taken from the 
NASA Extragalactic 
Database (NED; \url{http://nedwww.ipac.caltech.edu}), where there are currently 
275 references for individual distance measurements to the LMC}.  

\subsection{{\it Spitzer} Infrared Data}

{\it Spitzer} imaging of the LMC was carried out as part of the Surveying the 
Agents of a Galaxy's Evolution \citep[SAGE;][]{mm06} legacy program.  
The observations covered an area of $\sim 7^\circ \times 7^\circ$ over the LMC 
in all IRAC (3.6, 4.5, 5.8, and 8~$\mu$m) and MIPS (24, 70, and 160~$\mu$m) 
bandpasses.
The individual observations were calibrated, combined, and mosaicked by the 
SAGE team, using a custom pipeline \citep[see][for further details]{mm06}.  
Because we are only interested in diffuse emission, we use point-source 
subtracted versions of these images prepared by the SAGE team (see right 
panel of Figure~\ref{fig-1}).  
For the present study, we only make use of the MIPS 24, 70, and 160~$\mu$m 
data, which have native resolutions of $\approx5$\farcs7, 17\arcsec, 
and 38\arcsec, respectively. 
The photometric uncertainties are conservatively taken to be 5, 10, 
and 15\% at 24, 70, and 160~$\mu$m, respectively (see the MIPS Instrument 
Handbook\footnote[2]{\url{http://irsa.ipac.caltech.edu/data/SPITZER/docs/mips/mipsinstrumenthandbook/}}).  

The majority of the emission at 24~$\mu$m arises from dust heated in the 
vicinity of massive stars, which are the progenitors of core-collapse 
supernovae. 
Consequently, the 24~$\mu$m data are used as a proxy for the CR 
source distribution in our phenomenological modeling.
We use the full infrared data (i.e., 24, 70, and 160~$\mu$m imaging) to 
estimate the radiation field energy density over the LMC and 
30~Dor (see $\S$\ref{sec-imsm}). 

\begin{figure*}[th]
\begin{center}
\epsscale{1.17}
\plottwo{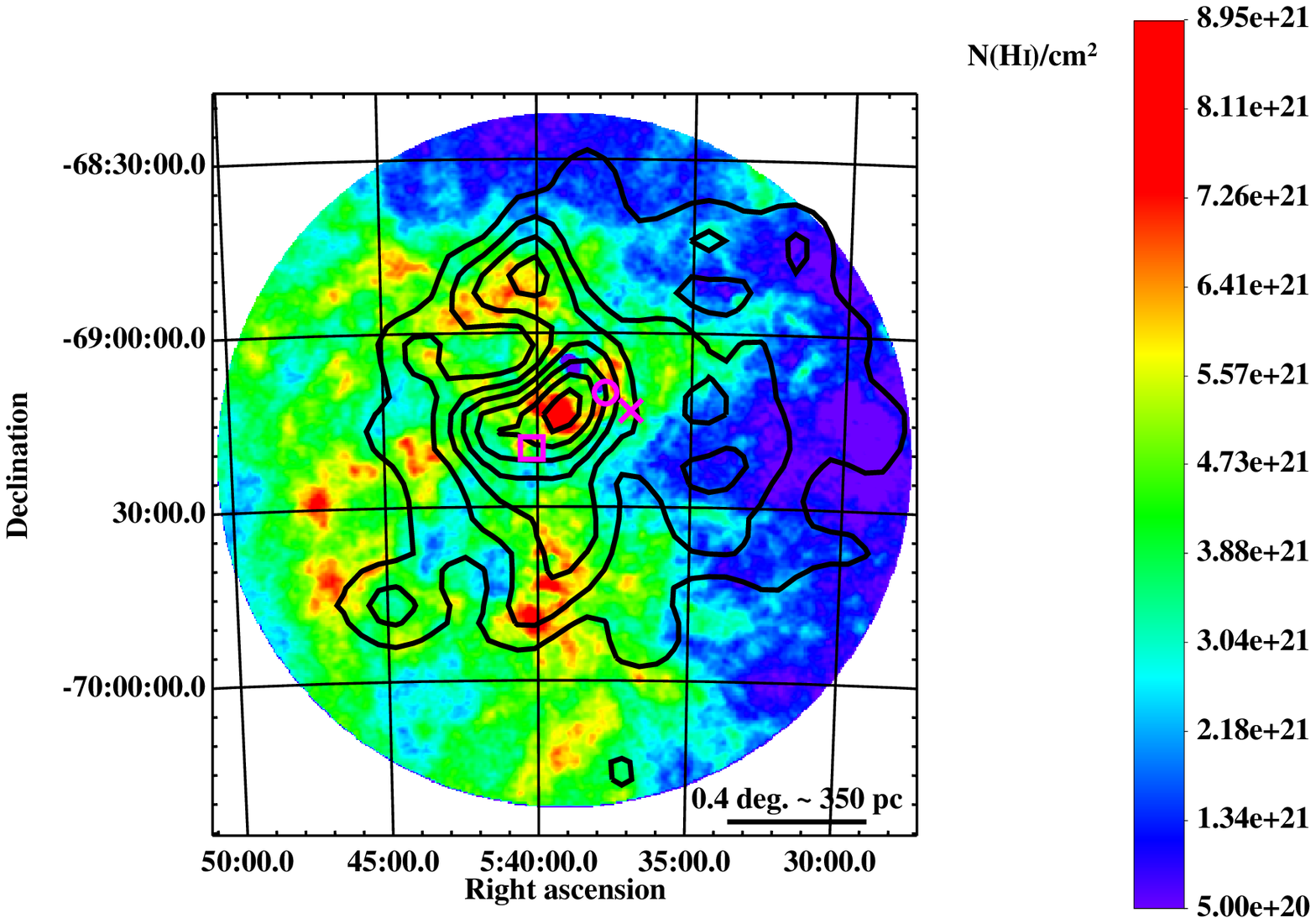}{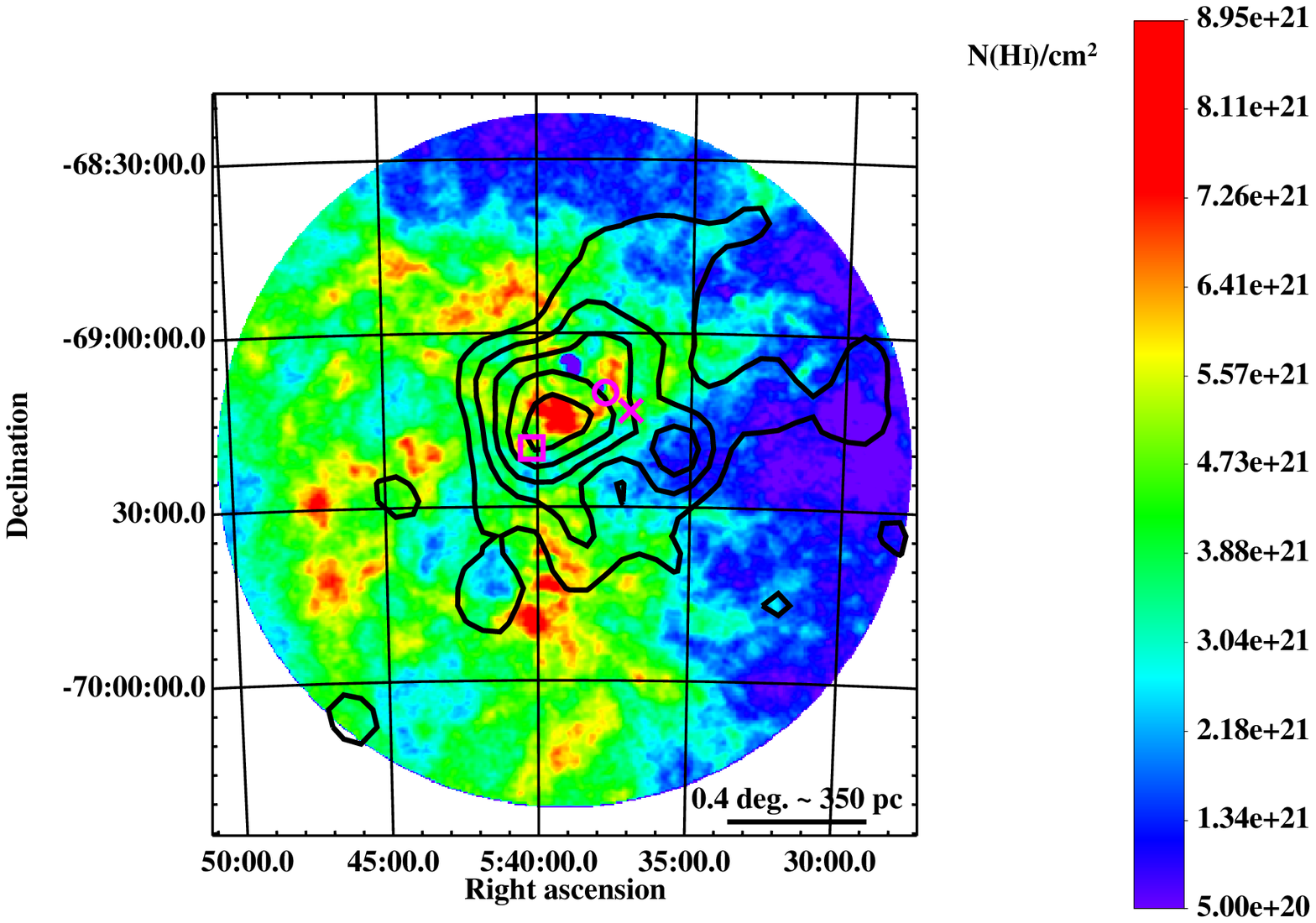}
\end{center}
\caption{
The same region on the sky as shown in Figure~\ref{fig-1}, but now displaying the H{\sc i} column density map \citep{lss03,sk03}.  
In the left panel, residual $1-3$~GeV \gray{} contours are overlaid on the H{\sc i} column density map with the same intervals as shown in Figure~\ref{fig-1}.  
In the right panel, residual $3-10$~GeV \gray{} contours are overlaid on the H{\sc i} column density map starting at $0.2\times10^{6}$~counts~sr$^{-1}$ with linear intervals of $0.178\times10^{-6}$~counts~sr$^{-1}$.    
Also shown are the two Crab-like pulsars, PSR~J0540$-$6919 (square) and PSR~J0537$-$6910 (circle), 
along with a known background X-ray source RX~J0536.9$-$6913 (`X') that 
are coincident with the line-of-sight towards 30~Dor.  
\label{fig-2}}
\end{figure*}

\subsection{Parkes+ATCA Radio Data}
\label{sec-datrad}
We use combined 1.4~GHz 
single-dish (Parkes 64~m) 
and interferometric (ATCA) data presented in \citet{ah06}, where a much 
more detailed description of the data preparation can be found.  
We employ a version of the radio 
map that has point sources removed (see left panel of Figure~\ref{fig-1})  
from the map using a median filter 
technique \citep[see][]{ah06} because more than 90\% of such sources are 
expected to be background AGN \citep{mdm97}.  
The angular resolution of the final combined mosaic is 40\arcsec, and 
sets the resolution for our analysis between the radio and infrared data.  
At our assumed distance for the LMC, this projects to a linear 
scale of $\approx10$~pc.  

Because we are interested in the non-thermal component of the radio 
data, we subtract the contribution from thermal (free-free) emission using 
a scaled version of the 24~$\mu$m image following Equation~2 of \cite{ejm06b}.  
This relation is based on the empirical correlation found between 
the 24~$\mu$m and extinction-corrected Pa$\alpha$ luminosities from 
star-forming regions within NGC~5194 by \citet{dc05}.  
Although not universal \citep[e.g.,][]{ppg06,dc07}, it has proved to be a 
good first-order estimate of the thermal radio emission in nearby galaxies 
when compared to thermal fractions derived from multifrequency 
radio data \citep[e.g.,][]{ejm08}.  

To check the reliability of our free-free emission estimate, we 
compare with the thermal fraction derived from standard radio spectral 
decompositions using single-dish radio data from
the literature:
1.4~GHz \citep{uk89}, 2.3~GHz \citep{pim87}, and 2.45~GHz \citep{rfh91}.  
There is a known discrepancy between the total flux from the 1.4~GHz 
single-dish map and the combined Parkes$+$ATCA map being used here, 
as \citet{ah07} report a global 1.4~GHz flux density that is 1.3 times 
smaller than the single dish estimate from \citet{uk89}.  
This difference is largely attributed to an underestimation of the beam 
width for the single-dish data.  
Because the single-dish data were reduced and analyzed self-consistently 
by \citet{rfh91}, we assume that each map is affected in a similar way.  

Using these multifrequency radio data, and assuming a constant non-thermal 
spectral index of $\alpha_{\rm NT} \approx 0.7$ \citep{rfh91}, we derive a thermal fraction at 1~GHz that is $\approx1.2$ times larger than the value of 25\% found using the 24~$\mu$m map over our region of interest surrounding 30~Dor shown in Figure~\ref{fig-1}.    
This difference is consistent with results from a high-resolution radio 
continuum survey of M~33 \citep{fat07b}, where a standard 
radio spectral decomposition was found to yield 
thermal fractions for 11 H{\sc ii} 
complexes that were overestimated by an average factor of 1.5 at 8.3~GHz.
However, even if we scale the 24~$\mu$m-derived free-free map by a 
factor of 1.2 to match the total thermal fraction measured by the radio 
spectral decomposition in our region of interest, our results are not 
significantly affected (see \S\ref{sec-smresults}).  

\subsubsection{CR Electron Energy Estimate}
\label{sec-datradcre}
We estimate the typical energies of the CR electrons emitting the 
observed non-thermal 1.4~GHz emission. 
For electrons propagating with a pitch angle $\theta$ in a magnetic field 
of strength $B$ with isotropically distributed velocities, such 
that $<\sin^{2}\theta> = \twothirds$ leading to $B_{\perp} \approx 0.82 B$, 
then a CR electron emitting at a critical frequency $\nu_{c}$ will have 
an energy
 
\begin{equation}
\label{eq-nuBE}
\left(\frac{E_e}{\rm GeV}\right) = 8.8 \left(\frac{\nu_c}{\rm GHz}\right)^{1/2} \left(\frac{B}{\rm \mu G}\right)^{-1/2}.  
\end{equation}

\noindent
Integrating the free-free corrected 1.4~GHz flux over the region shown in 
Figure~\ref{fig-1} ($S^{\rm NT}_{\rm 1.4GHz} = 102.26$~Jy; 
see \S\ref{sec-imsm}), we use 
the revised minimum energy calculation of \citet{bk05}, assuming a 
proton-to-electron number density ratio of $K_{0} \approx 100 \pm 50$, a 
non-thermal spectral index of $\alpha_{\rm NT} \approx 0.7 \pm 0.1$,  and a 
path length of $l \approx 1 \pm 0.5$\,kpc, to calculate a minimum energy 
magnetic field strength of $\sim 11 \pm 4$~$\mu$G.  
Using this range of values in Equation~\ref{eq-nuBE}, our 1.4~GHz 
maps are most sensitive to $\sim 3\pm0.5$~GeV CR electrons.  
If the minimum energy condition does not hold, and we instead assume an 
extreme range in $B$, such as $3-50$~$\mu$G, the corresponding CR 
electron energies will range from $6-1.5$~GeV.
However, the (in-)validity of the minimum energy assumption is uncertain 
and often assumed (as we do in this paper).
We briefly discuss the impact of this on the derived properties of the 
diffusion coefficient in \S\ref{cr-trans}.  

\subsection{{\it Fermi}-LAT Data} 
\label{sec-fermdat}
The \fermilat{} instrument, event reconstruction, and response are 
described in \citet{wat09}.
In this paper, we use
events and instrument response functions (IRFs) for 
the standard low-background ``Clean'' events 
corresponding to the Pass~7 event selections\footnote[3]{We use the P7V6\_CLEAN event class in this work; see \url{http://fermi.gsfc.nasa.gov/ssc/data/analysis/documentation/Pass7\_usage.html}}.  
To minimize the contribution from the very bright Earth limb, 
we restrict the event selection and exposure calculation to zenith 
angles $<100^\circ$.  
We selected all Front-converting events, for which
the point-spread function (PSF) is narrowest, within a 20$^\circ$ square 
region of interest (RoI) centered 
on RA~$05^{\rm h} 38^{\rm m} 42^{\rm s}$, 
Dec.~$-69^\circ 06\arcmin 03\arcsec$ (J2000) for $\approx32$ months 
of sky survey data from 2008 August 4 until 2011 March 24.
Exposure maps and the PSF for the pointing 
history of the observations were generated using the standard 
\fermilat{} ScienceTools package available from the \fermi{} Science 
Support Center\footnote[4]{\url{http://fermi.gsfc.nasa.gov/ssc/data/analysis/}}.
The exposure of the instrument over the RoI for the data taking period used
in this analysis is very uniform.

We constructed foreground-subtracted \gray{} maps over the same region 
using the standard \fermilat\ diffuse 
emission model\footnote[5]{\url{http://fermi.gsfc.nasa.gov/ssc/data/access/lat/BackgroundModels.html}} 
together with the appropriate isotropic background model and \gray{} point sources from the {\it Fermi}-LAT Second Source Catalog \citep{2FGL}, 
forward folding the combination with the exposure and PSF to obtain the 
total counts expected for these foregrounds.
These were subtracted from the data and the residual emission for the 
$1-3$~GeV energy interval is shown in 
Figure~\ref{fig-1} overlaid on the free-free corrected 
1.4~GHz and 24~$\mu$m maps.
The effective 68\% containment radius for the \fermilat\ PSF over this energy bin for the event selection used is $\approx 0.50^\circ$ ($\approx 440$ pc), 
assuming an $E_\gamma ^{-2.1}$ spectrum for the \gray{} emission, which is a sufficient approximation to the spectral
shape determined by \cite{aa10-lmc} for the LMC \gray{} emission over this
energy range\footnote[6]{Weighting with an $E_\gamma ^{-2.8}$ spectrum gives 
a marginally different result, being larger by $\approx$3.5\%.}  
Note that the effective full width at half-maximum (FWHM) of the \fermilat\ PSF over this energy bin is $\approx 0.38^\circ$ ($\approx 330$ pc), significantly smaller than the 68\% containment radius quoted above.  

\begin{deluxetable}{ccccc}
\tablecaption{Effective \gray{} and Proton Energies for Different \gray{} Energy Ranges  \label{tab:protonE}}
\tablehead{
\colhead{Proton Energy} & \multicolumn{4}{c}{\gray{} Energy Range} \\
\colhead{Index} & \multicolumn{2}{c}{$1-3$~GeV} & \multicolumn{2}{c}{$3-10$~GeV}\\
\colhead{} & \colhead{$E_{\gamma}^{\rm eff}$} & \colhead{$E_{p}^{\rm eff}$} & \colhead{$E_{\gamma}^{\rm eff}$} & \colhead{$E_{p}^{\rm eff}$}\\
\colhead{} & \colhead{(GeV)}& \colhead{(GeV)}& \colhead{(GeV)}& \colhead{(GeV)}
}
\startdata
  2.1  &       1.63  &       22.4  &       5.10  &       69.2\\
  2.2  &       1.61  &       19.5  &       5.04  &       60.4\\
  2.3  &       1.60  &       17.2  &       4.99  &       53.6\\
  2.4  &       1.58  &       15.4  &       4.93  &       48.1\\
  2.5  &       1.60  &       14.1  &       4.87  &       43.6\\
  2.6  &       1.55  &       12.5  &       4.82  &       39.8\\
  2.7  &       1.54  &       11.5  &       4.77  &       36.6\\
  2.8  &       1.53  &       10.5  &       4.72  &       33.9
 \enddata
\end{deluxetable}

Similarly, we constructed a map for the $3-10$~GeV residual emission.  
This is shown in Figure~\ref{fig-2} where we overlay the $1-3$ 
and $3-10$~GeV contours on the \hi\ column density 
data \citep{lss03,sk03}.  
The effective 68\% containment radius for this map is $\approx 0.24^\circ$ ($\approx 210$ pc), 
with an effective FWHM of $\approx 0.19^\circ$ ($\approx 170$ pc),
again assuming an $E_\gamma ^{-2.1}$ spectrum for the \gray{} emission over this
energy bin.
However, the $3-10$~GeV map has much lower statistics, 
even near 30~Dor, contrasting with the $1-3$~GeV \gray{} map, which 
has $\ga3$ times more events.  
Therefore, we base the majority of our spatial 
analysis on the $1-3$~GeV map, and only use the higher energy \gray{}
data for testing the energy dependence of the CR 
propagation (see $\S$\ref{sec-3-10GeV}).

\subsubsection{CR Nuclei Energy Estimate}

As for electrons and the 1.4~GHz emission, we estimate the typical 
energies of the CR nuclei emitting $\gamma$ rays for the energy ranges that we consider in this paper.  
For non-AGN dominated star-forming galaxies like the Milky Way and the LMC, 
the assumption that the CR nuclei are the predominant relativistic particle
population producing the \gray{} emission in the \fermilat\ 
energy range is reasonable \citep{Strong2010}.

We define the function

\begin{equation}
\begin{split}
G(E_\gamma,E_p) &= \left(\int_{E_p}^\infty dE_p' \, \beta \frac{dJ_p}{dE_p'}
\frac{d\sigma_{pp}(E_\gamma,E_p')}{dE_\gamma}\right) \bigg / \\
&		\left(\int_{E_p^{\rm min}(E_\gamma)}^\infty dE_p' \, \beta \frac{dJ_p}{dE_p'}
		\frac{	d\sigma_{pp}(E_\gamma,E_p')}{dE_\gamma}\right) - \frac12,
\end{split}
\end{equation}

\noindent
where $E_p$ and $E_\gamma$ are the proton and \gray{} energies, 
$\beta=v/c$, $dJ_p/dE_p' \propto E_p'^{-\alpha}$ is the CR proton 
flux, $d\sigma_{pp}/dE_\gamma$ is the differential cross section 
for \gray{} production calculated as in \citet{ms98},  
and $E_p^{\rm min}$ is the minimum proton energy required to produce 
a photon of energy $E_\gamma$.  
The effective proton energy, $E_p^{\rm eff}$, for a given $E_\gamma$ is given by the root of $G(E_{\gamma}, E_p)$ so that contributions to the \gray{} flux at $E_\gamma$ below and above $E_p^{\rm eff}$ are equal. 
The effective proton energy is a function of the proton spectral 
index $E_p^{\rm eff}=E_p^{\rm eff}(\alpha)$, which we show for different 
$\alpha$ for the $1-3$~GeV and $3-10$~GeV energy ranges in Table~\ref{tab:protonE}.
We also show the effective (average) \gray{} energy for the $1-3$~GeV and $3-10$~GeV 
energy ranges, calculated assuming the same spectral index for the $\gamma$ rays as for the protons.
This assumption is valid for thin-target interactions by 
the CR nuclei with gas, which applies for the ISM surrounding 
30~Dor and the greater LMC.  

The effective proton energy changes by a factor of $\sim2$ when
the power-law index of the proton spectrum changes from 2.1 to 2.8, while 
the effective energy of $\gamma$ rays does not change appreciably. 
For the \gray{} energies, this is not surprising because
the effective energy is calculated over a relatively narrow energy bin. 
The effective proton energy change is a combination of the fairly steep 
rise in the inclusive pion production cross section as the momentum
of the interacting proton increases, approximately $\propto E^{0.5}$ 
between 10 and 100~GeV \citep[see, e.g., Figure~2a from][]{1986A&A...157..223D}.
When the proton spectrum is steep (e.g., with an index $\sim2.8$), this 
rise is not large enough to compensate for the steeply falling number of 
high-energy protons. 
When the proton spectrum is flat (e.g., with an index $\sim2.1$)
the rise in the inclusive cross section increases the contribution of 
high-energy protons, making the $E_p^{\rm eff}$ significantly larger.  

\subsection{Image Registration and Smearing Analysis}
\label{sec-imsm}
All images were cropped to a common field-of-view and regridded to a 
common pixel scale.  
To properly compare each image at the same resolution, maps were convolved with
an appropriate PSF.  
For the radio and infrared (24, 70, and 160~$\mu$m) data, we convolve all 
images to the resolution of the 1.4~GHz image, which has a FWHM of 40\arcsec, 
using a Gaussian beam having a FWHM equal to the quadrature difference 
between the final and original FWHM values of each image.    

To match the resolution of the 24~$\mu$m data to that of the \gray{} maps, we convolve the 24~$\mu$m image (at its native resolution) with the in-flight \fermilat\ PSF, described in Section \ref{sec-fermdat}.  
Because the FWHM of the 24~$\mu$m PSF is $<1$\% of the effective FWHM over the \gray{} energy bins used in our analysis, we do not attempt to correct for this additional broadening.

Next, we apply our image smearing analysis (described below) to the resolution-matched 
images to determine the extension of the non-thermal radio and \gray{} 
morphologies relative to the 24~$\mu$m morphology.  
The procedure largely follows that described in \citet{ejm06b,ejm08}, where 
we refer the reader for a more detailed description.  
Our underlying assumption is that the only difference between these 
distributions is due to the diffusion and energy losses of the CRs 
because of the common origin for the CR leptons and nuclei. 
We focus on the 24~$\mu$m data in the smearing analysis rather than 
the 70~$\mu$m data, as was done by \citet[e.g.,][]{ejm08}, because it 
traces warmer dust and has been found to be more peaked near star-forming 
regions than the 70~$\mu$m morphology \citep{gxh04,ejm06a}.  
For instance, typical 24$\mu$m knots surrounding 30~Dor tend to be $\sim$10~pc in size.  
Thus, the 24~$\mu$m maps likely act as a better source function for 
recently accelerated CRs. 

To summarize the procedure, we convolve the \emph{entire} 24~$\mu$m image 
of the LMC by a parameterized kernel $\kappa({\bf r})$ and compute the 
residuals between the free-free corrected 1.4~GHz and smoothed 24~$\mu$m 
maps.    
The smoothing kernel is a function of a two-dimensional position 
vector \textbf{\textit{r}}, having a magnitude of $r =  (x^{2} + y^{2})^{1/2}$, 
where $x$ and $y$ are the right ascension and declination offsets on the 
sky, respectively.  
We investigate both exponential and Gaussian kernels because a preference of 
kernel type may suggest different CR transport effects. 
The exponential and Gaussian kernels have the forms of 
\(\kappa(\textbf{\textit{r}})= e^{-\textbf{\textit{r}}/l}\) and 
\(\kappa(\textbf{\textit{r}})= e^{-\textbf{\textit{r}}^{2}/l^{2}}\), respectively, 
where $l$ is the $e$-folding scale length.  
Thus, for the Gaussian kernel, $\sigma^{2} = l^{2}/2$.  

Exponential kernels, which have broader tails than Gaussian kernels of the 
same scale length, imply energy loss and/or escape timescales less than, 
or comparable to, the diffusion timescale.  
Gaussian kernels, on the other hand, suggest a simple random walk scenario.  
Because the LMC lacks a well-definied disk, and has a low 
inclination \citep[$i \approx 31^\circ$;][]{sn04}, 
we assume spherical symmetry for 30~Dor.  
Hence, there is no need to modify the position vector by geometric factors. 

We calculate the normalized-squared residuals between the 
radio and the infrared images 
 
\begin{equation}
\label{eq-phi}
\phi(l) = \frac{\Sigma[Q^{-1}\tilde{I}_{j}(l) - R_{j}]^2}{\Sigma R_{j}^{2}}
\end{equation}

\noindent
where $R$ is the radio image with free-free emission removed, $\tilde{I}(l)$ 
is the infrared image smoothed by a kernel with scale 
length $l$,  \(Q = \Sigma I_{j}/\Sigma R_{j}\) is used as a normalization 
factor (i.e., the global 24~$\mu$m/1.4~GHz ratio), 
and the subscript $j$ indexes each pixel.  
Only pixels detected with a significance of $>3\sigma$ of the RMS noise in 
each map are used in the calculation of the residuals.    
The minimum in $\phi$ defines the best-fit scale length, which is taken 
to be the typical distance traveled by the CR electrons.  
Similarly, we perform the same procedure using the 24~$\mu$m 
and $1-3$~GeV \gray{} maps to determine the typical distance traveled by 
the CR nuclei.  

For this procedure, generically, if there is no broadening due to propagation effects, the minimum of Equation \ref{eq-phi} will be at $l = 0$, because both the radio and \gray{} data have already reduced to the same angular resolution.
Below, we obtain kernel scale lengths indicating that significant smoothing of the 24~$\mu$m image is required to improve the match with the radio and \gray{} images.  
While the scale lengths obtained for the radio/24~$\mu$m residual analysis are much larger than the angular resolution of these maps, for the \gray{}/24~$\mu$m data, the derived scale lengths are comparable to the angular resolution after convolution with the \fermilat\ PSF.  
The technique is capable of detecting scale lengths that are a small fraction of the resolution (i.e., $\ll \Delta l = 50$~pc, which is the scale length step size used in the analysis), and is more sensitive for higher signal-to-noise ratio maps.
We emphasize that the kernel scale lengths obtained in this case are in addition to the smoothing of the 24~$\mu$m data to match the \fermilat\ PSF.
If there was no additional effect from propagation, the derived scale lengths would be zero, whereas we detect a meaningfully non-zero scale length.  


The uncertainty in $\phi$ is estimated by numerically propagating the 
uncertainties in the input images as measured by the 1$\sigma$ RMS noise of 
each map, with the uncertainty for the best-fit scale length then estimated 
as the range in scale length corresponding to that 
from $\min(\phi)$ to $\min(\phi) + {\rm unc}(\phi)$ along the residual curve.  
We note that this places a lower limit on the uncertainty for the best-fit scale lengths since there is additional uncertainty on the 1.4~GHz thermal fraction estimation and foreground subtraction for the \gray{} maps.  
The calculation of the residuals and photometry was carried out within 
an aperture having a radius of $1^\circ$ ($\approx875$~pc), 
encompassing 30~Dor, centered 
at $05^{\rm h}\,39^{\rm m}\,07^{\rm s}, -69\degr 22\arcmin\, 02\arcsec$ (J2000, 
see Figure~\ref{fig-1}).  

We use the 24, 70, and 160~$\mu$m photometry to calculate the total 
infrared (IR, $8-1000$ $\mu$m) luminosity over this region by fitting these 
data to the spectral energy distribution (SED) models of \citet{dh02} and 
integrating the best-fit SED between 8 and 1000~$\mu$m.  
The individual 24, 70, and 160\,$\mu$m flux densities are $0.42\pm0.02$, $4.1\pm0.41$, and $8.0\pm1.7$ $\times10^4$~Jy, respectively.  
In addition to photometric uncertainties, the above errors include a term for the mean RMS noise values in the convolved 24, 70, and 160~$\mu$m maps, which are measured to be $\sim$0.045, 0.685, and 13.2~MJy~sr$^{-1}$, respectively.  
We obtain an IR luminosity 
of \(L_{\rm IR} = 1.15 \pm 0.12 \times 10^{42}\,{\rm erg\,s^{-1}}\)  
($3.00 \pm 0.31 \times 10^{8}\,L_{\sun}$).  
Taking this value, we estimate the corresponding radiation field 
energy density

\begin{equation}
\label{eq-urad}
U_{\rm rad} \approx \frac{2\pi}{c}I_{\rm bol} \ga \frac{L_{\rm IR}}{2 A_{\rm IR}c}\left(1 + \sqrt{\frac{3.8\times10^{42}}{L_{\rm IR}}}\right), 
\end{equation}

\noindent
where $I_{\rm bol}$ is the bolometric surface brightness, $c$ is the speed 
of light, and $A_{\rm IR} \approx 2.4\,{\rm kpc^{2}}$ is the area over which 
the photometry was measured. 
All quantities in Equation~\ref{eq-urad} are in cgs units.    
This calculation is for radiation emitted near the surface of a 
semitransparent body, and the parenthetical term provides an empirically 
derived correction for non-absorbed UV emission \citep{efb03}, resulting 
in a value of $U_{\rm rad} \approx 2.39 \pm 0.17 \times 10^{-12}\,{\rm erg\,cm^{-3}}$ ($1.49 \pm 0.11\,{\rm eV\,cm^{-3}}$) averaged over the volume considered.

\begin{figure}
\scalebox{1.1}{
\plotone{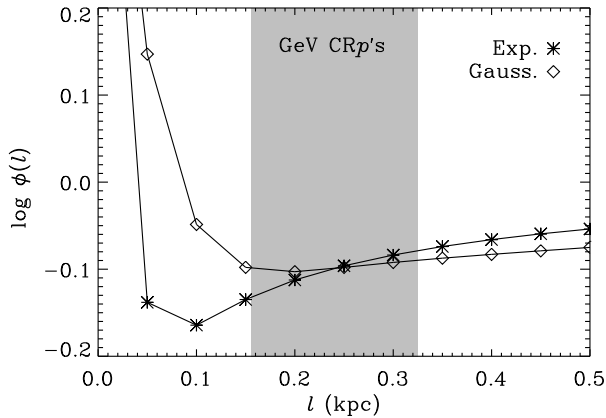}}
\caption{
The residuals between the free-free corrected 1.4~GHz and 
smoothed 24~$\mu$m (as defined in \S \ref{sec-imsm}) plotted as a function 
of both exponential (asterisks) and Gaussian (diamonds) of kernel scale 
lengths.  
The shaded region indicates the extrapolated propagation length for GeV 
protons based on the compact morphology of the \gray{} emission observed 
by the \fermilat\ \citep{aa10-lmc}.
The value reported by \citet{aa10-lmc} is the $\sigma$ from a modeled 
Gaussian. 
The values for the Gaussian scale lengths agree within errors.  
\label{fig-3}}
\end{figure}

\begin{figure}
\scalebox{1.1}{
\plotone{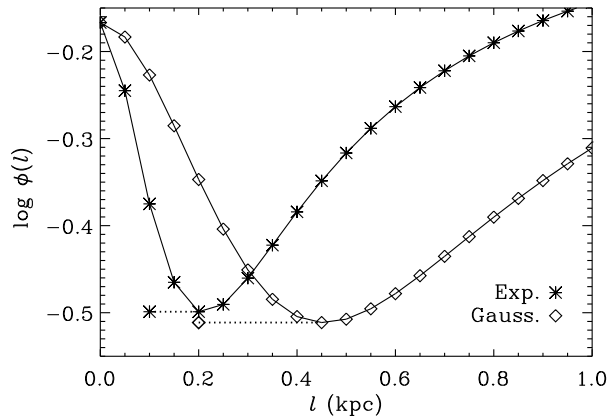}}
\caption{
The residuals between the $1-3$~GeV \gray{} and smoothed 24~$\mu$m 
images (as defined in \S \ref{sec-imsm}) plotted as a 
function of both exponential (asterisks) and Gaussian (diamonds) of kernel 
scale lengths.  
The asterisk and diamond attached to the horizontal dotted lines 
indicate the location of the best-fit scale length measured for the CR 
electrons using exponential and Gaussian kernels (see Figure~\ref{fig-3}).  
\label{fig-4}}
\end{figure}

\section{Results}
We present our results from comparing the morphologies of 30~Dor as measured 
by warm dust emission at 24~$\mu$m, synchrotron emission at 1.4~GHz, and 
\gray{} emission at $1-3$~GeV.  
We compare the morphologies both qualitatively, through a visual 
inspection, as well as quantitatively using the image smearing analysis 
described above.  

\subsection{24~$\mu$m, 1.4~GHz, and $1-3$~GeV \gray{} Morphologies}
\label{sec-morph}
Figure~\ref{fig-1} shows the residual $1-3$~GeV \gray{} emission overplotted 
as contours on the free-free corrected 1.4~GHz radio and 24~$\mu$m warm dust emission in 
the left and right panels, respectively.  
While there is a general correspondence between the \gray{}, radio, and 
infrared emission, the peak of the \gray{} emission is clearly offset 
from the peak of the radio and infrared emission, which appear to be 
nearly co-spatial.  
This lack of correspondence between the peak of the \gray{} emission with 
the radio and infrared emission leads one to question whether the 
observed \gray{} emission is indeed associated with 30~Dor.  

In \citet{aa10-lmc}, it was noted that \gray{} emission 
near 30~Dor may have a non-negligible contribution from two 
Crab-like pulsars, PSR~J0540$-$6919 ($05^{\rm h}\,40^{\rm m}\,11.2^{\rm s}, -69\degr 19\arcmin\, 54\arcsec$; J2000) 
and PSR~J0537$-$6910 ($05^{\rm h}\,37^{\rm m}\,47.4^{\rm s}, -69\degr 10\arcmin\, 20\arcsec$; J2000).  
In addition, there is a background X-ray source (RX~J0536.9$-$6913: $05^{\rm h}\,36^{\rm m}\,57.8^{\rm s}, -69\degr 13\arcmin\, 26\arcsec$; J2000) that 
is also coincident with the line-of-sight toward 30~Dor,  
which could also contribute to the observed emission.   
The locations of these point sources are shown in Figure~\ref{fig-1} 
and Figure~\ref{fig-2}, and are near, but not at, the peak of 
the $1-3$ and $3-10$~GeV \gray{} emission.  
It is also interesting to see that the $1-3$ and $3-10$ \gray{} contours 
appear to peak close to a region of high H{\sc i} column density in 
Figure~\ref{fig-2}.  
It is possible to subtract fitted \gray{} point sources at the the locations of the pulsars and AGN, resulting in a situation where 30~Dor does not exhibit \gray{} emission in excess of the subtracted foreground model. 
The implications of such a scenario are described in $\S$4.1.
However, we note that 
there is no detection of pulsed \gray{} emission so far reported 
from either of the pulsars.
Also, no variability is detected from the region during our period of 
analysis.
If RX~J0536.9$-$6913 is a \gray{} emitting AGN, the lack of variability makes it
difficult to determine if this candidate source
contributes significantly in our RoI.  

\subsection{Results from the Smearing Analysis}
\label{sec-smresults}
In Figure~\ref{fig-3} we plot the residuals ($\phi$, see \S \ref{sec-imsm}) 
between the free-free corrected 1.4~GHz and smoothed 24~$\mu$m maps as a 
function of exponential and Gaussian kernel scale lengths.  
The residuals between the free-free corrected 1.4~GHz and 24~$\mu$m maps 
are decreased by more than a factor of $\sim$2 after smoothing the 24~$\mu$m 
map using either exponential or Gaussian kernels.  
A slightly larger improvement is found by using an exponential kernel, 
suggesting energy losses and/or escape may be important for the CR 
electrons.     
The corresponding best-fit exponential and Gaussian kernel scale lengths are 
100$^{+10}_{-10}$ and 200$^{+100}_{-55}$~pc, respectively.  
We note that the estimated $\sim$3~GeV CR electron propagation length 
reported here assuming random walk diffusion (i.e., using a Gaussian kernel) 
is consistent within errors to the GeV CR proton confinement length reported 
by \citet{aa10-lmc}, whose estimate assumed Gaussian 
profiles (see Figure~\ref{fig-3}).  
\citet{aa10-lmc} report the 
Gaussian $\sigma$ to be $\sigma = 170 \pm 60$~pc, so 
we multiply this number by $\sqrt{2}$ for proper comparison with the scale 
length definition for our Gaussian kernel.  

As discussed in \S\ref{sec-datrad}, we test how our results are 
affected by assuming that the thermal fraction of the 1.4~GHz radio 
map is being 
underestimated by the 24~$\mu$m maps relative to what is derived using 
single-dish radio data at 1.4, 2.3, and 2.45~GHz.  
By scaling the 24~$\mu$m-derived free-free maps by a factor of 1.2 to match 
the total thermal fraction measured by the radio spectral decomposition, and 
repeating the smearing analysis, we find that the best-fit exponential 
scale length is still 100~pc, while the best-fit Gaussian scale length is 
slightly increased to 250~pc.  
A larger improvement is again found using an exponential kernel relative to 
a Gaussian kernel.  
The results therefore do not appear to be significantly affected by increasing 
the thermal fraction estimate to match that from the radio spectral 
decomposition method.  

Similarly, in Figure~\ref{fig-4}, we plot the residuals between 
the $1-3$~GeV \gray{} and smoothed 24~$\mu$m maps as a function of 
exponential and Gaussian kernel scale lengths.   
The residuals between the $1-3$~GeV and 24~$\mu$m maps are decreased by a 
factor of $\sim$2 after smoothing the 24~$\mu$m map using 
either exponential or Gaussian kernels.  
While a slightly larger improvement is found by using exponential kernels 
in the comparison between the 1.4~GHz and 24~$\mu$m maps, we do not 
find a preference in kernel type for the $1-3$~GeV \gray{} map.  
The corresponding best-fit exponential and Gaussian kernel scale lengths 
are 200$^{+40}_{-10}$ and 450$^{+65}_{-50}$~pc, respectively.  
These best-fit scale lengths are significantly larger (i.e., a factor 
of $\sim$2) than what was measured for the 1.4~GHz maps.  
Because the 1.4~GHz and $1-3$~GeV \gray{} maps probe different energy CR 
particle populations, the corresponding differences in their best-fit scale 
lengths may arise from different diffusion speeds.  
This scenario is discussed in $\S$\ref{sec-crdiff}.  
 
\section{Discussion}
Using a phenomenological image smearing model, we have estimated the 
typical propagation lengths of $\sim3$~GeV CR electrons and $\sim20$~GeV CR 
nuclei by comparing the spatial distributions of smoothed 24~$\mu$m maps 
images with (free-free corrected) 1.4~GHz 
and (foreground subtracted) $1-3$~GeV \gray{} maps, respectively.  
From the analysis, there is a difference between the typical 
distances travelled by $\sim3$~GeV CR electrons and $\sim20$~GeV protons 
(assuming a proton energy index of $2.1$), where 
the 20~GeV CR protons are found to travel $\sim$2 times further, on average, 
independent of kernel type.  
However, it is additionally worth including a brief discussion of the 
physical implications for the case where 30~Dor may not be emitting 
in $\gamma$ rays.  

\subsection{The Case for a Dark 30~Dor in $\gamma$ rays}
As stated in $\S$3.1, there are both pulsars and a background X-ray emitting 
source that may contribute to the observed \gray{} emission towards 30~Dor.  
For this case, the situation has 30~Dor emitting at 1.4~GHz, therefore 
containing diffusing CR electrons, but no observational signature for the
CR nuclei.  
Explaining this would require the CR nuclei to escape the system without 
interacting with the interstellar gas.  

Taking the above estimate for the bolometric luminosity from 30~Dor 
(i.e., the IR luminosity given in \S\ref{sec-imsm} corrected by the 
parenthetical term in 
Equation~\ref{eq-urad} for non-absorbed UV emission) together with the 
updated star formation rate (SFR) calibrations given in \citet{ejm11b}, we 
estimate a corresponding SFR of $\sim$0.15~$M_{\sun}\,{\rm yr}^{-1}$.  
This value is consistent with others in the 
literature \citep[e.g.,][]{ah07, hz09, bl10}.
Assuming this value and the sensitivity of the \fermilat\ data used in our
analysis, the absence of \gray{} emission from 30~Dor 
clearly disagrees 
with the empirical correlations between SFR or the product of supernova 
rate and the gas mass 
with \gray{} luminosity found for the local group and nearby 
starbursts \citep[e.g.,][]{aa10-Fg-sfr,lw11}.
Furthermore, theoretical expectations of such scaling relations and their 
implications for the contribution by normal star-forming galaxies 
to the diffuse \gray{} extragalactic background \citep[e.g.,][]{pf01,bdf10} 
would then be questionable.  
However, because there is currently no strong evidence linking the \gray{} 
emission from 30~Dor to the known point sources in the field, we discuss the 
physical interpretation of our analysis with the assumption that it arises 
from 30~Dor.  

\subsection{Cosmic Ray Propagation from 30~Dor}
\label{sec-crdiff}
To interpret our findings on the kernel scale lengths, we must consider 
the dependence of gas density on the synchrotron and \gray{} emission.
The diffuse $\gamma$ rays in the energy range that we have used ($1-3$~GeV) are 
predominantly from CR nuclei interacting with the interstellar 
gas, and therefore $S_{\gamma} \propto n_{p} n_{\rm ISM}$ where $n_{p}$ and 
$n_{\rm ISM}$ are the densities of the CR nuclei and gas, respectively.
The diffuse synchrotron emission is proportional to the CR electron density,
$n_e$, and the square of the magnetic field 
strength, $B$, i.e., $S_{\rm sync} \propto n_{e} B^2$.
Assuming flux-freezing scaling of 
$B \propto \sqrt{n_{\rm ISM}}$ \citep[e.g.][]{rss88, nb97,rc99}, the 
synchrotron emission becomes proportional to the product of the electron 
density and the density of the interstellar 
gas, $S_{\rm sync} \propto n_{e} n_{\rm ISM}$. 
This simple scaling argument therefore suggests that the differences in 
appearance of the \gray{} and radio images are due to differences in 
the distributions of the CR nuclei and electrons.
This was tested by normalizing the \gray{} and free-free corrected radio 
images by the \hi\ column density map before applying the smearing analysis, 
recovering results consistent with those described in \S\ref{sec-smresults}.  

\subsubsection{Cosmic Ray Transport Properties} 
\label{cr-trans}
We assume that the transport of the CRs can be described by a simple 
random-walk process with an rigidity-dependent spatial diffusion 
coefficient $D_{R}$ \citep[e.g.,][]{ginz80}.  
The diffusion coefficient is defined as 

\begin{equation}
\label{eq-DE}
D_{R} = D_{0} \left(\frac{R}{\rm GV}\right)^{\delta} \approx \frac{l^{2}_{\rm diff}}{\tau_{\rm diff}}, 
\end{equation}

\noindent
where $D_{0}$ is the normalization constant 
and $l_{\rm diff}$ is the characteristic distance that CRs travel after a 
time $\tau_{\rm diff}$.  
For our Gaussian kernels, we can relate $l_{\rm diff}$ to the corresponding 
best-fit scale lengths such that $l_{\rm diff}^{2} = \sigma^{2} = l^{2}/2$.  

If we assume that the CR electrons and nuclei are injected by the 
same source(s) and have been propagating through the ISM for the same 
length of time, we find that the $\sim20$~GeV CR nuclei diffusion 
coefficient is $\sim4$ times larger than that for the $\sim3$~GeV CR 
electrons.  
Solving Equation~\ref{eq-DE} for $\delta$, we find that the 
diffusion coefficient scales as $(R/{\rm GV})^{0.69 \pm 0.15}$ from the 
exponential best-fit scale lengths and as $(R/{\rm GV})^{0.81 \pm 0.30}$ using 
the Gaussian best-fit scale lengths.  
Errors on $\delta$ were estimated by a standard Monte Carlo approach using 
the uncertainties in the best-fit scale lengths\footnote[7]{
$\delta$ was calculated by taking 1000 random samples, 
picking best-fit scale lengths from a 
normal distribution having a dispersion set by their lower and upper-bound uncertainties.  
The standard deviation of this set is taken as the uncertainty on $\delta$.}.  
If the minimum energy assumption does not hold, and we instead 
consider the extreme range of $B \approx 3-50~\mu$G described earlier 
in \S\ref{sec-datradcre}, we find corresponding $\delta$ values 
of $\approx 0.5 - 1.2$.  

These \emph{model-independent} values of $\delta$ are consistent with the  
value of $\sim0.6-0.7$ 
used in empirical diffusion models to fit the observed secondary-to-primary 
ratios, typically boron-to-carbon (B/C), for the Milky Way.
In addition, our derived values for $\delta$ generally exceed those for physically motivated turbulence theories, being larger than the nominal value of $\onethird$ for Kolomogorov \citep{ak41}, and marginally consistent with the value of $\onehalf$ for Iroshnikov-Kraichnan \citep{psi64,rhk65} turbulence spectra.        
However, the values for the Milky Way are found for a large volume in a 
spiral galaxy, while the results of our analysis are for a single, highly 
active star-forming region: 
30~Dor exhibits a complex network of kinematic features 
including slow ($v \la 100\,{\rm km\,s^{-1}}$) and 
fast ($100\,{\rm km\,s^{-1}} \la v \la 300\,{\rm km\,s^{-1}}$) expanding 
shells powered by stellar winds from young massive stars and 
supernovae \citep{chuken94}.  

Because of strong radiative energy losses as they propagate, the corresponding diffusion lengths for the $\sim3$~GeV CR 
electrons may be underestimated by our best-fit scale lengths, which would 
result in an overestimate for $\delta$.  
This is suggested by the fact that an exponential kernel resulted in a lower 
minimum residual 
between the 1.4~GHz and 24~$\mu$m maps than a Gaussian kernel.
However, as discussed below, 
the additional cooling of electrons may not significantly affect our estimate 
for $\delta$.    

Taking the average derived value for $\delta$ (i.e., 0.75), together 
with the age of the current star formation activity responsible for 
supernova remnants accelerating the CRs, we can estimate the diffusion 
coefficient normalization factor $D_{0}$.  
The 30~Dor complex is known to contain many non-coeval stellar 
populations ranging in age from $<1-10$~Myr from {\it HST} 
spectroscopy \citep{wb97}.  
Recently, using a Bayesian analysis, \citet{jrm11} fit {\it Spitzer}-IRS 
data for 30~Dor using mid-infrared SED models (continuum + lines), arriving 
at a luminosity-weighted age for the system of $\sim3$~Myr.  
This suggests that CRs associated with this star-forming event have 
been accelerated in the supernova remnants of very massive O-stars, and that 
the bulk of CRs have yet to be accelerated by the supernova remnants from 
the more numerous and less massive (i.e., $\sim 8 M_{\sun}$) stars with 
lifetimes of $\sim30$~Myr.  
Note that this age is less than the estimated cooling lifetime 
$\tau_{\rm cool} \sim 12\pm2.3$~Myr for the 1.4~GHz emitting CR electrons 
as they propagate through the ISM of 30~Dor\footnote[8]{This maximum 
lifetime is estimated using the combined 
energy losses due to synchrotron, inverse Compton, bremsstrahlung, and 
ionization processes \citep[see, e.g.,][]{ejm09c}.  
Using the previously estimated values for the minimum energy magnetic 
field strength ($B_{\rm min} \approx 11\pm4\,\mu$G; see $\S$2.2), the radiation 
field energy density ($U_{\rm rad} \approx 2.39 \pm 0.17 \times 10^{-12}\,{\rm erg\,cm^{-3}}$; see $\S$2.4), and assuming an average 
ISM density of $n_{\rm ISM} \approx 2 \pm 1\,$cm$^{-3}$ \citep{sk03}, we 
estimate a cooling time for the observed 1.4~GHz emitting ($\sim3$~GeV) 
electrons of $\sim$$12\pm2.3$~Myr.  
We note that the individual lifetimes estimated against synchrotron, 
inverse Compton, 
bremsstrahlung, and ionization losses are 33, 55, 43, and 140~Myr, 
respectively}.  

Using this mean age of 3~Myr in Equation~\ref{eq-DE}, together with the 
best-fit Gaussian scale lengths corresponding 
to $l_{\rm diff} \approx 140\,$pc and 320~pc for the 3~GeV and 20~GeV 
electrons and nuclei, respectively, we find a diffusion coefficient 
normalization constant 
of $D_{0} \approx 0.9-1.0 \times 10^{27}\,{\rm cm^{2}\,s^{-1}}$.  
This is more than an order of magnitude lower than those found in the Milky Way 
\citep[e.g., $\approx 5\times10^{28}$ cm$^2$ s$^{-1}$,][]{apj11-trotta-crprop}.  
However, analytical solutions for the perpendicular diffusion coefficient (i.e., diffusion across magnetic field lines) are 
found to be much smaller. 
For example, \citet{sbls10} report a perpendicular diffusion coefficient of 
$D_{\perp} \approx 3.0-30 \times 10^{26}\,{\rm cm^{2}\,s^{-1}}$.  
Because 30~Dor is an active star-forming region, which likely has a strong 
turbulent magnetic field and lacks a large-scale regular field 
like the Galaxy, it may not be surprising that our derived 
value for the diffusion coefficient normalization factor is intermediate 
in the range for the perpendicular diffusion coefficients obtained by 
other authors.  

\subsubsection{Results Including $3-10$~GeV Maps}
\label{sec-3-10GeV}
So far we have only discussed the results comparing the propagation 
for $\sim3$~GeV CR electrons and $\sim20$~GeV CR nuclei.  
While it seems that the cooling time for the 1.4~GHz ($\sim$3~GeV) CR 
electrons is a factor of $\sim4$ times larger than the average age of the 
stellar population in 30~Dor, trying to interpret the differences in their 
propagation lengths is complicated by the fact that the $\sim 3$~GeV CR 
electrons lose energy more rapidly than the $\sim 20$~GeV CR nuclei.  
That the electron energy losses are important is suggested by the 
result that the residuals between the 1.4~GHz and 24~$\mu$m maps are 
decreased more when using exponential kernels rather than Gaussian kernels.  
We therefore include the $3-10$~GeV \gray{} map to provide an additional 
estimate 
for the propagation length of CR nuclei because it is sensitive to much 
higher energy CR nuclei (i.e., $\sim70$~GeV for a proton energy index of 
$2.1$).  
Employing these data, we can try to determine whether the difference between 
the CR electrons 
and protons is in fact the result of rigidity-dependent diffusion.    

Applying the same smearing analysis to the 1.4~GHz radio, $1-3$~GeV, 
and $3-10$~GeV \gray{} maps, but limiting the area of the residual calculation 
to that where there are $3-10$~GeV \gray{} events (i.e., $\approx \onehalf$ 
the area having $1-3$~GeV \gray{} events), we obtain the following results.  
The best-fit exponential and Gaussian scale lengths are 50 and 150~pc, 
respectively, for $\sim3$~GeV electrons, 150 and 300~pc, respectively, 
for $\sim20$~GeV CR protons, and 200 and 400~pc, respectively, 
for $\sim70$~GeV protons.
Because the detected $3-10$~GeV events are concentrated around the 
center of 30~Dor, lacking a significant extended component as observed 
in the $1-3$~\gray{} map, 
it is not 
surprising that the scale lengths obtained are generally smaller than those calculated from the residuals over a larger area.  

Using all six possible combinations of these values to solve for 
$\delta$ results in a median and dispersion of $\delta \approx 0.66 \pm 0.23$.  
This is consistent with the energy dependence estimated above using 
the 1.4~GHz radio and $1-3$~GeV \gray{} maps. 
We note that the value of $\delta$ estimated by comparing only the \gray{} maps, which is independent of assumptions for the magnetic field strength and proton energy index, 
is $\approx0.51$, whereas the median when only including the 1.4~GHz maps is $\approx0.79$, suggesting that the best-fit scale lengths for the CR electrons may be underestimated due to additional energy losses.  
This is consistent with exponential kernels working slightly better to tighten 
the correlation between the 1.4~GHz and 24~$\mu$m maps compared to Gaussian kernels.  
Additionally, the estimate for the value of $\delta$ using only the \gray{} maps is independent of assumptions for the magnetic field strength
However, we emphasize that the $3-10$~GeV \gray{} map is 
statistically limited, having $\ga3$ times fewer events than the $1-3$~GeV maps.

\begin{figure}
\scalebox{1.1}{
\plotone{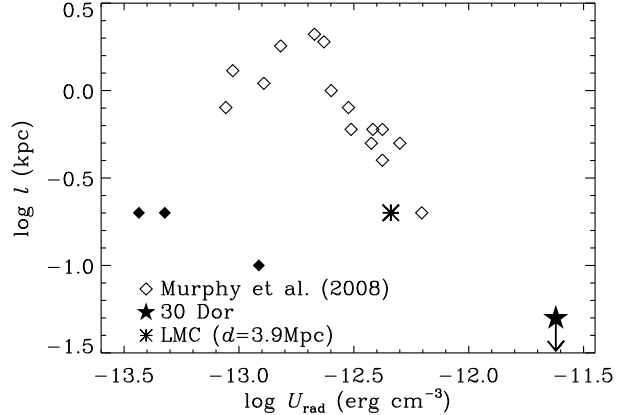}}
\caption{
Best-fit exponential scale lengths plotted against 
radiation field energy density (measured by the infrared surface brightness) 
for the galaxies studied in \citet[][diamonds]{ejm08}, the LMC (asterisk), 
and 30~Dor (filled star).
The open diamonds indicate spirals, while the filled diamonds show the 
location of dwarf irregular galaxies included in the \citet{ejm08} study.  
The location of 30~Dor (filled star) follows the trend of decreasing 
exponential scale length, indicating shorter propagation lengths of CR 
electrons with increasing radiation field energy density for the 
more active star-forming disk galaxies in the \citet{ejm08} study.   
We plot the CR electron propagation distance for 30~Dor as an upper 
limit (see \S \ref{sec-sf}.)
For the dwarf irregular galaxies, the short scale lengths are thought to be 
due to an increase in the escape of the CR electrons from the systems.
If we set the LMC at the mean distance of the \citet{ejm08} dwarf 
irregulars (i.e., $\sim3.9$~Mpc), its propagation scale length is more 
consistent with the \citet{ejm08} spirals.  
\label{fig-5}}
\end{figure}

\begin{figure}
\scalebox{1.1}{
\plotone{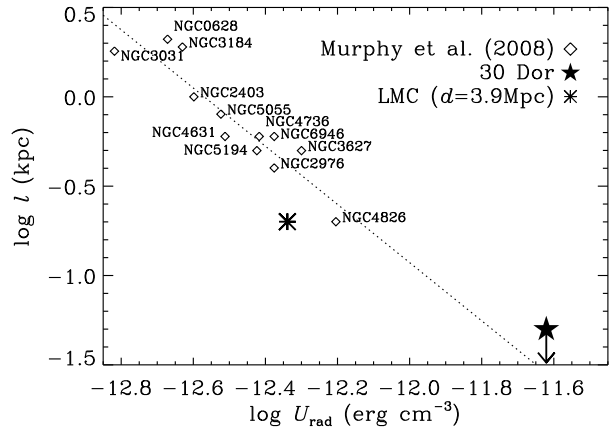}}
\caption{
The same as Figure~\ref{fig-5}, except that we show only galaxies having high disk-averaged star formation rates.   
A distinct trend of decreasing CR electron propagation length with increasing radiation field energy 
density can be seen.  
Galaxies from the \citet{ejm08} study have been labeled.  
The dotted line is an ordinary least squares fit {\it only} to the galaxies from \citet{ejm08}.  
The locations of 30~Dor (filled star, as an upper limit) and the entire LMC (asterisk; scaled to $d=3.9$~Mpc) are plotted.  
\label{fig-6}}
\end{figure}

\subsubsection{Propagation Length vs. Star Formation Activity}
\label{sec-sf}
Having estimated the average distance traveled by CR electrons for 
an individual star-forming region, 30~Dor, it is interesting to see 
how these results compare with similar estimates of CR electron propagation 
distances for entire galaxies.  
\citet{ejm06b,ejm08} reported a correlation between the 
CR electron propagation distance and galaxy surface brightness such that 
CR electrons are found to travel shorter distances, on average, in galaxies 
having higher star formation activity.  

In Figure~\ref{fig-5} we plot the results 
from \citet[][i.e., the best-fit exponential scale lengths projected in the 
plane of the sky versus $U_{\rm rad}$]{ejm08} along with our estimates 
for 30~Dor.  
We note that various systematics and selection effects (e.g., distance, 
inclination) were rigorously investigated by \citet{ejm08}.  
They additionally tested the differences between isotropic kernels and 
kernels projected in the plane of each galaxy disk, finding a 
preference, i.e., smaller residuals, for isotropic kernels.  

The $U_{\rm rad}$ estimate for 30~Dor was calculated from the surface 
brightness measured over an area of 2.4~${\rm kpc^{2}}$, the same region for 
which the residuals were estimated (see \S \ref{sec-imsm}).   
For this case, the best-fit scale length was measured by comparing the 
free-free corrected 1.4~GHz with smoothed 70~$\mu$m maps to allow a proper 
comparison with the results of \citet{ejm08}.  
The best-fit exponential and Gaussian kernel scale lengths were 0 and 50~pc, 
respectively.  
Because \citet{ejm08} plot exponential kernel scale lengths, we plot the 
result from smoothing the 70~$\mu$m map of 30~Dor with a Gaussian kernel as 
an upper limit.  

Included in Figure~\ref{fig-5} are both spirals and star-forming irregulars.
The latter have markedly small best-fit scale lengths for their 
values of $U_{\rm rad}$ relative to the sample of spirals.  
This is thought to be caused by increased escape of CR electrons in 
these systems \citep[e.g.,][]{jmc05,jmc06,ejm08}.  
To see how the behavior of entire galaxies compares with the LMC as 
a whole, we repeat the image smearing analysis using the 70~$\mu$m maps 
after first projecting the LMC to the mean distance of the \citet{ejm08} 
sample irregulars, $d=3.9$~Mpc to take possible resolution 
effects into account.  
We calculate the residuals and $U_{\rm rad}$ within an aperture having 
a radius of $\approx3.1$~kpc (2\farcm75 at the distance of 3.9~Mpc) 
centered at $05^{\rm h}\,18^{\rm m}\,50^{\rm s}, -68\degr 50\arcmin\, 49\arcsec$ 
(J2000).  
This is approximately the area over which the 24~$\mu$m emission was 
detected at $\ga3\sigma$ level.    
The best-fit exponential and Gaussian scale lengths are 200 and 400~pc, 
respectively.  
This places the LMC close to the main trend among the star-forming spirals.  

Focusing only on the region containing the star-forming spirals, we 
re-plot the disk galaxies along with 30~Dor and the LMC in 
Figure~\ref{fig-6}.  
The dotted-line is a least-squares fit to the \citet{ejm08} galaxy disks, 
excluding the LMC and 30~Dor.  
Clearly, the position of 30~Dor in the plot is consistent with the
empirical trend describing spiral galaxies, but extrapolated to a surface 
brightness greater by almost an order of magnitude.  
Thus, this scaling relation appears to operate on the scales of entire 
galaxies all the way down to individual star-forming regions.   

\section{Conclusions}
Using a phenomenological image smearing model, we have estimated 
the typical propagation length of $\sim3$~GeV CR electrons and $\sim20$~GeV 
CR nuclei by comparing the spatial distributions of smoothed 24~$\mu$m 
and (free-free corrected) 1.4~GHz and $1-3$~GeV \gray{} maps, respectively.  
Below we list the major results and conclusions:

\begin{enumerate}
\item 
We estimate the typical distances traveled by $\sim$3~GeV CR electrons 
from 30~Dor to be $\sim 100-140$~pc.   
This is factor of $\sim$2 smaller than that estimated for $\sim$20~GeV CR 
nuclei, which is found to be $\sim 200-320$~pc.  

\item 
In our image-smearing analysis, we find that exponential kernels work 
slightly better to tighten the correlation between the 1.4~GHz and 24~$\mu$m 
maps compared to Gaussian kernels.  
In contrast, both exponential and Gaussian kernels are found to work equally 
well to tighten to correlation between the $1-3$~GeV \gray{} 
and 24~$\mu$m maps.  
This difference suggests that, unlike the CR nuclei, CR leptons suffer 
additional energy losses as they propagate through the ISM near 30~Dor on 
timescales less than, or comparable to, the diffusion timescale.  

\item
Assuming that the CR electrons and nuclei are produced by the same sources,
and that their propagation is well described by a random walk, the 
differences in their estimated propagation lengths suggest differences in 
their associated spatial diffusion coefficients.  
This allows us to make the first model-independent measurement of the 
energy dependence of the diffusion coefficient for an external galaxy.   
For CRs produced in this star-forming region, 
the $\sim$20~GeV CR proton diffusion coefficient is $\sim$4 times larger 
than that for $\sim$3~GeV CR electrons, scaling as $(R/{\rm GV})^{\delta}$ 
where $\delta \approx 0.7-0.8$.  
This value is consistent with that obtained by including the more statistically limited $3-10$~GeV 
\gray{} map ($\sim70$~GeV CR protons; $\delta \approx 0.66 \pm 0.23$).  

\item
The value of $\delta$ reported here is larger than the spectral index of the diffusion coefficient assuming Kolomogorov turbulence
and Milky Way secondary-to-primary ratios, and marginally consistent with Iroshnikov-Kraichnan turbulence.  
This may reflect the fact that 30~Dor region exhibits complex 
kinematic features, and fast expanding shells, resulting in a larger 
value of $\delta$.  

\item
Assuming that the CRs in 30~Dor are as old as the average stellar population ($\sim$3~Myr), we estimate a diffusion coefficient normalization 
constant of $D_{0} \approx 0.9-1.0 \times 10^{27}\,{\rm cm^{2}\,s^{-1}}$.  
This value is less than and similar to model-dependent estimates of the 
parallel and perpendicular diffusion coefficient for the Galaxy, respectively.  
The similarity between our estimate and perpendicular diffusion coefficient 
values may be due the 30~Dor region having a magnetic field configuration 
that is highly turbulent.  

\item 
We place our results for 30~Dor, and the LMC as a whole, in the context of 
the scaling relation between the typical CR electron propagation length and 
disk-averaged star formation activity for entire disk galaxies, where CR 
electron propagation is found to decrease with increasing star formation 
activity.  
This relation appears to apply to the LMC and the individual star-forming region of 30~Dor.  

\end{enumerate}

\acknowledgements

E.\,J.\,M. would like to thank Annie Hughes for both providing us with radio maps and for useful discussions which greatly helped shape the paper.  
We are grateful to the SAGE team for producing high quality data sets used in this study. 
T.~A.~P. and E.~J.~M. acknowledge support via NASA grant NNX10AE78G.
I.~V.~M. acknowledges support via NASA grant NNX09AC15G.

The \fermilat\ Collaboration acknowledges generous ongoing support from a number of agencies and institutes that have supported both the development and the operation of the LAT as well as scientific data analysis.  
These include the National Aeronautics and Space Administration and theDepartment of Energy in the United States, the Commissariat \`a l'Energie Atomique and the Centre National de la Recherche Scientifique / Institut National de Physique Nucl\'eaire et de Physique des Particules in France, the Agenzia Spaziale Italiana and the Istituto Nazionale di Fisica Nucleare in Italy, the Ministry of Education, Culture, Sports, Science and Technology (MEXT), High Energy Accelerator Research
Organization (KEK) and Japan Aerospace Exploration Agency (JAXA) in Japan, and the K.~A.~Wallenberg Foundation, the Swedish Research Council and the Swedish National Space Board in Sweden.

Additional support for science analysis during the operations phase is 
gratefully
acknowledged from the Istituto Nazionale di Astrofisica in Italy and the 
Centre National d'\'Etudes Spatiales in France.

This work is based in part on observations made with the {\it Spitzer} Space 
Telescope, which is operated by the Jet Propulsion Laboratory, California 
Institute of Technology under a contract with NASA.
This research has additionally made use of the NASA/IPAC Extragalactic 
Database (NED), which is also operated by the Jet Propulsion Laboratory, California 
Institute of Technology, under contract with the NASA.

\bibliographystyle{apj}
\bibliography{lmc_crprop}

\end{document}